\documentclass[]{emulateapj}
\usepackage{graphicx}

\graphicspath{{Figures/}}

\makeatletter
\newcommand{\mion}[2]{#1$\;${\sc\@roman{#2}}}
\newcommand{\cion}[2]{\mbox{\scriptsize#1$\;${\sc\@roman{#2}}\relax}}
\newcommand{\ionl}[3]{\mbox{#1$\;${\small\rmfamily\@Roman{#2}}$\,\lambda\lambda\,$#3\relax}}
\makeatother

\def\pg0953{\mbox{PG~0953$+$415}\relax}

\newcommand{\zabs}[1]{\mbox{$z_{\rm abs} = #1$}\relax}

\def\kms{\mbox{km~s$^{-1}$}\relax}

\slugcomment{Submitted for publication in {\it The Astrophysical Journal Letters}}

\begin{document}

\title{The \ion{O}{6} Absorbers Toward PG0953+415: \\ High
Metallicity, Cosmic-Web Gas Far From Luminous
Galaxies\altaffilmark{1}}

\altaffiltext{1}{Based on observations with (1) the NASA/ESA {\it
Hubble Space Telescope}, obtained at the Space Telescope Science
Institute, which is operated by the Association of Universities for
Research in Astronomy, Inc., under NASA contract NAS 5-26555, and (2)
the NASA-CNES/ESA {\it Far Ultraoviolet Spectroscopic Explorer}
mission, operated by Johns Hopkins University, supported by NASA
contract NAS 5-32985, and (3) the Gemini Observatory, which is
operated by the AURA, Inc., under a cooperative agreement with the NSF
on behalf of the Gemini partnership: the NSF (United States), the
PPARC (United Kingdom), the NRC (Canada), CONICYT (Chile), the ARC
(Australia), CNPq (Brazil), and CONICET (Argentina)}

\author{Todd M. Tripp,\altaffilmark{2} Bastien Aracil,\altaffilmark{2}
 David V. Bowen,\altaffilmark{3} and Edward
 B. Jenkins\altaffilmark{3}}

\altaffiltext{2}{Department of Astronomy, University of Massachusetts,
710 N. Pleasant St., Amherst, MA 01003-9305;
aracil@fcrao1.astro.umass.edu}

\altaffiltext{3}{Princeton University Observatory, Peyton Hall,
Princeton, NJ 08544}

\begin{abstract}
The spectrum of the low-redshift QSO PG0953+415 ($z_{\rm QSO}$ =
0.234) shows two strong, intervening \ion{O}{6} absorption systems. To
study the nature of these absorbers, we have used the Gemini
Multiobject Spectrograph to conduct a deep spectroscopic galaxy
redshift survey in the $5' \times 5'$ field centered on the QSO. This
survey is fully complete for $r' < 19.7$ and is 73\% complete for $r'
< 21.0$. We find three galaxies at the redshift of the higher$-z$
\ion{O}{6} system ($z_{\rm abs}$ = 0.14232) including a galaxy at
projected distance $\rho = 155 h_{70}^{-1}$ kpc. We find no galaxies
in the Gemini field at the redshift of the lower$-z$ \ion{O}{6}
absorber ($z_{\rm abs}$ = 0.06807), which indicates that the nearest
galaxy is more than 195 $h_{70}^{-1}$ kpc away or has $L < 0.04
L^{*}$. Previous shallower surveys covering a larger field have shown
that the $z_{\rm abs}$ = 0.06807 \ion{O}{6} absorber is affiliated
with a group/filament of galaxies, but the nearest known galaxy has
$\rho =$ 736 $h_{70}^{-1}$ kpc.  The $z_{\rm abs}$ = 0.06807 absorber
is notable for several reasons. The absorption profiles reveal simple
kinematics indicative of quiescent material. The \ion{H}{1} line
widths and good alignment of the \ion{H}{1} and metal lines favor
photoionization and, moreover, the column density ratios imply a high
metallicity: [M/H] = $-0.3 \pm 0.12$.  The $z_{\rm abs} = 0.14232$
\ion{O}{6} system is more complex and less constrained but also
indicates a relatively high metallicity.  Using galaxy redshifts from
the {\it Sloan Digital Sky Survey (SDSS)}, we show that both of the
PG0953+415 \ion{O}{6} absorbers are located in large-scale filaments
of the cosmic web.  Evidently, some regions of the web filaments are
highly metal enriched. We discuss the origin of the high-metallicity
gas and suggest that the enrichment might have occurred long ago (at
high $z$).
\end{abstract}

 \maketitle

\section{Introduction}

The intergalactic medium (IGM) is an important frontier for studies of
galaxy evolution and cosmology for several reasons. The IGM is
expected to be the main repository of baryons at all
epochs, but the IGM physical characteristics are predicted to change
from predominantly cool ($T \sim 10^{4}$ K), photoionized gas at high
redshifts to predominantly shock-heated ``warm-hot'' gas ($T = 10^{5}
- 10^{7}$ K) at the current epoch (Cen \& Ostriker 1999; Dav\'{e} et
al. 1999). The exchange of matter and energy between galaxies and the
IGM, e.g., via gas accretion or supernova-driven ``feedback'', may play 
profound roles in the evolution of galaxies, groups, and clusters (Voit
2005).  Observations with the {\it Hubble} Space Telescope Imaging
Spectrograph (STIS) as well as the {\it Far Ultraviolet Spectroscopic
Explorer (FUSE)} have shown that \ion{O}{6} absorption lines are
frequently detected in the spectra of low-redshift QSOs (e.g., Tripp
et al. 2000; Danforth \& Shull 2005; Lehner et al. 2006).  In
collisional ionization equilibrium (CIE), \ion{O}{6} has its greatest
abundance at $T \approx 10^{5.5}$ K and therefore can, in principle,
be used to detect the warm-hot IGM and hot galactic outflows.
However, \ion{O}{6} can also arise in photoionized intergalactic gas
or gas that is not in ionization equilibrium, and only a few
\ion{O}{6} systems (e.g., Savage et al. 2005) show definitive evidence
of hot gas.

One means to investigate the nature and implications of \ion{O}{6}
absorbers is to study their relationships with nearby galaxies/galaxy
structures (e.g., Sembach et al. 2004; Bregman et al. 2004; Stocke et
al. 2006; Prochaska et al. 2006). For this purpose, we have obtained
spectroscopic redshifts of galaxies in the fields of several low$-z$
QSOs with the Gemini Multiobject Spectrograph (GMOS) on the
Gemini-North 8m telescope. In this {\it Letter} we report first
results from one of the fields observed with GMOS, that of the
well-studied QSO PG0953+415.

\section{Data}

\begin{figure*}
\centering
\scalebox{1.00}{\includegraphics{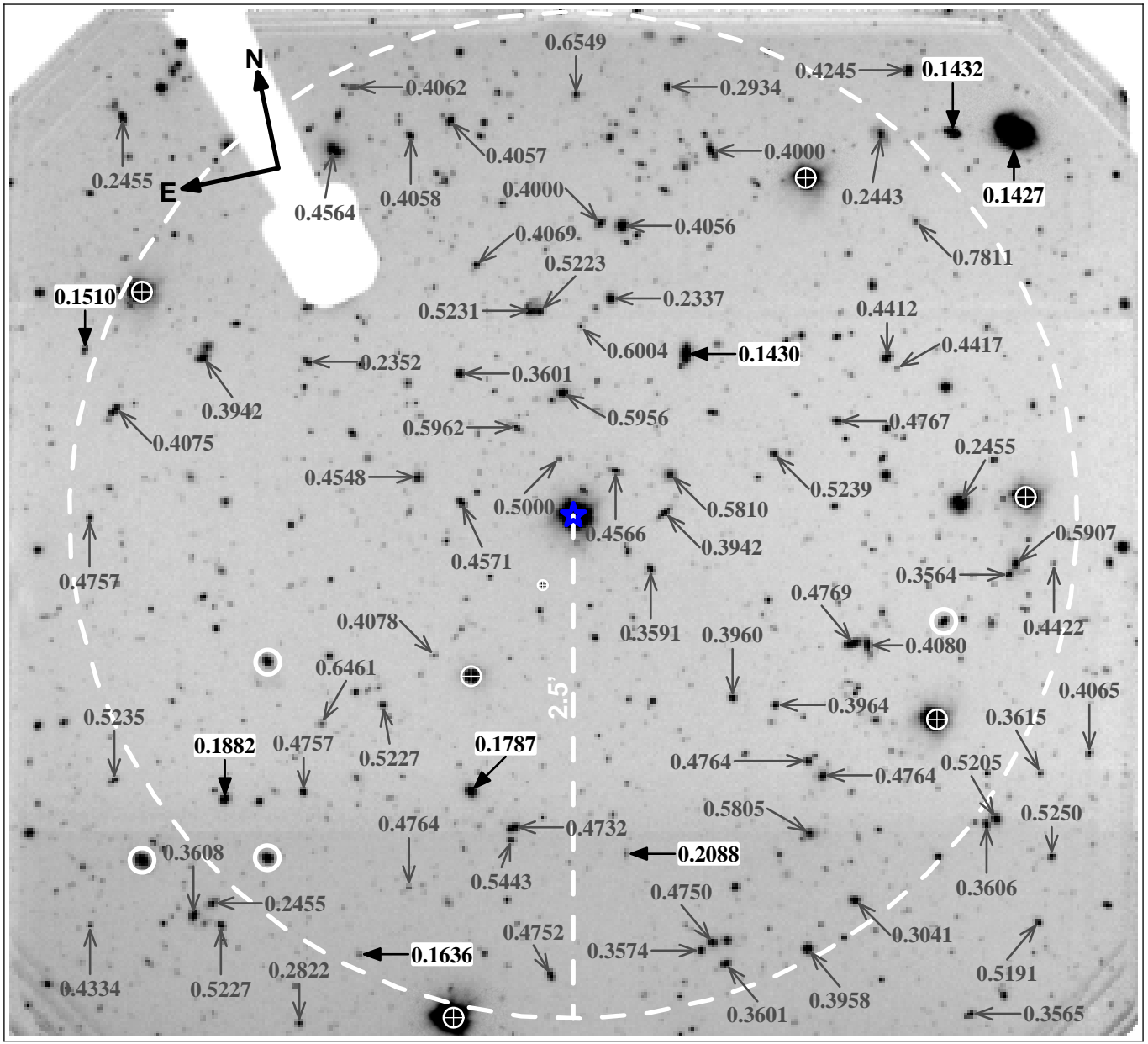}}
\caption{{\bf PLATE 1:} Optical $r'$ image of the PG0953+415 field
recorded with GMOS on the Gemini-North telescope. The QSO is in the
center, and the white dashed line shows a circle of $2\farcm 5$
radius. Galaxies are labeled with their spectroscopic redshifts if
known.  The spectroscopic redshifts place most of the galaxies behind
the QSO; redshifts in white boxes indicate foreground galaxies with
redshifts that are lower than $z_{\rm QSO}$ by at least 5000 \kms .
Bright stars are marked with crossed circles. The white region was
blocked by the GMOS guide star probe (which could not be positioned
outside of the science image). The four brightest galaxies without
spectroscopic redshifts are marked with small white
circles.}\label{fig:gmos_img}
\end{figure*}

\begin{figure*}
\centering
\scalebox{1.00}{\includegraphics{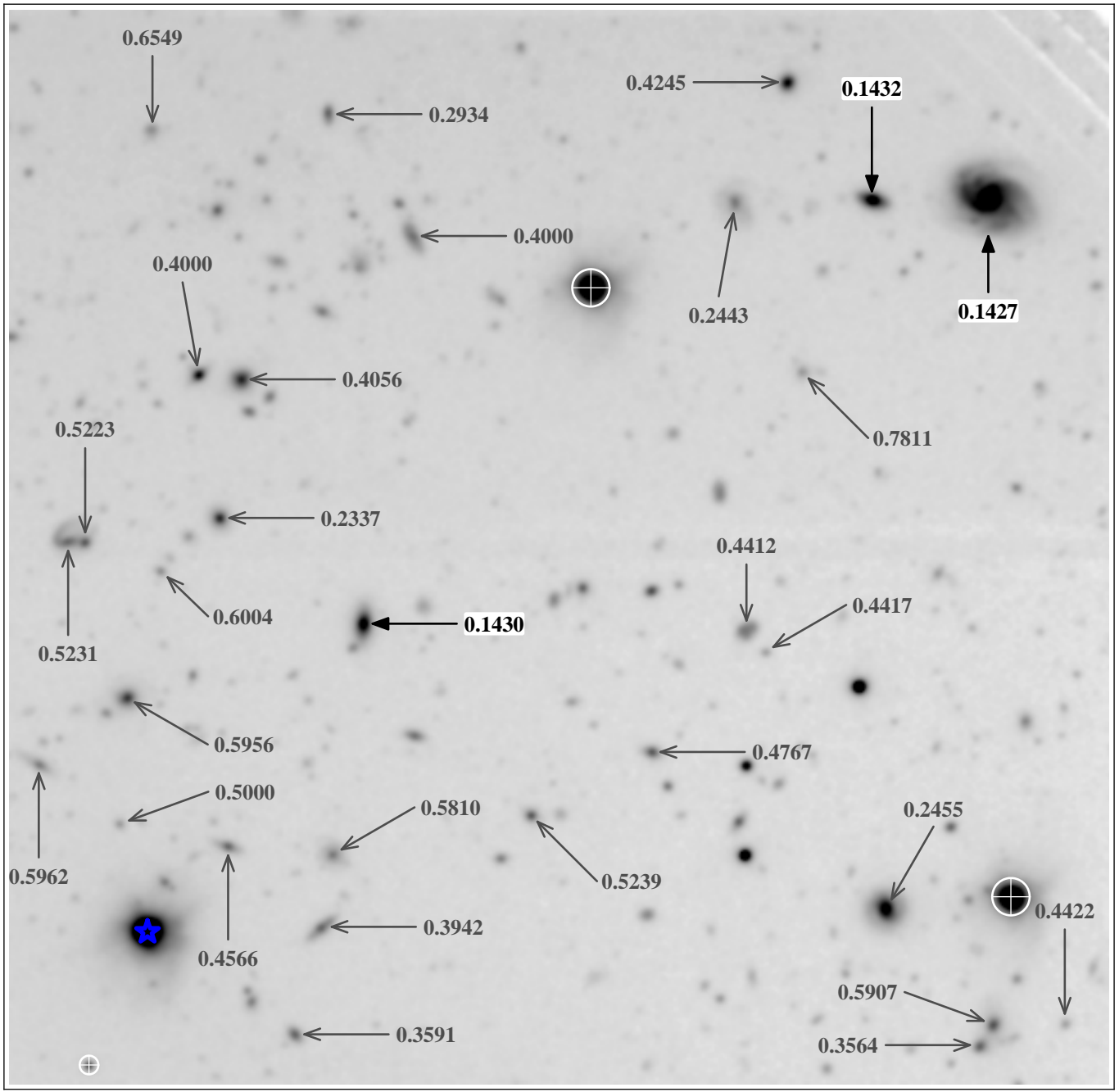}}
\caption{{\bf PLATE 2:} Zoom-in on the northwest portion of the GMOS
$r'$ image shown in Figure~\ref{fig:gmos_img}.  This figure shows the
faint galaxies in the immediate vicinity of the QSO (the point source
marked with a star at lower left) as well as the three galaxies at $z
= 0.143$ associated with the \ion{O}{6} absorber discussed by Tripp \&
Savage (2000). As in Figure~\ref{fig:gmos_img}, redshifts of galaxies
foreground to the QSO are marked in white boxes, and crossed circles
indicate stars.  The contrast in this image is adjusted to better show
the morphological structure of the foreground
galaxies.}\label{fig:gmos_zoom}
\end{figure*}

\begin{figure}
\centering
\scalebox{0.6}{\includegraphics{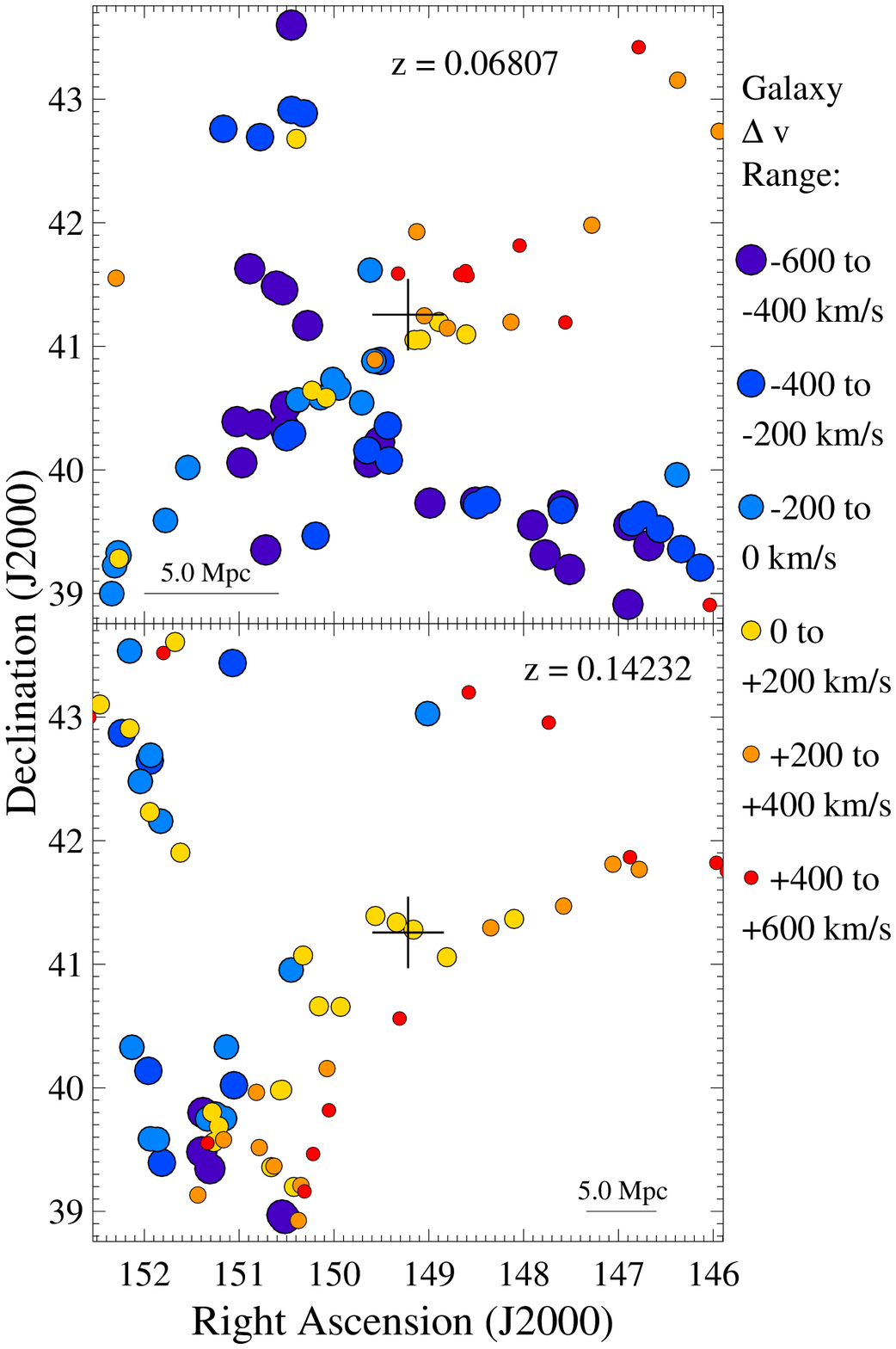}}
\caption{Distribution of galaxies in the vicinity of the \ion{O}{6}
absorbers at \zabs{0.06807} ({\it top}) and \zabs{0.14232} ({\it
bottom}) with spectroscopic galaxy redshifts from the {\it Sloan
Digital Sky Survey (SDSS)}. Only galaxies close to the absorber
redshifts are plotted, and the symbol size and color indicates the
difference between the galaxy redshift and the absorber redshift (in
km s$^{-1}$) as indicated in the key at right (e.g., the largest dark
blue symbols mark galaxies with $-600 < v_{\rm galaxy} - v_{\rm abs}
\leq -400$ km s$^{-1}$.  The direction of PG0953+415 is indicated with
a large plus symbol, and in each panel a projected distance of 5 Mpc
at the absorber redshift is indicated with a labeled horizontal black
line. The {\it SDSS} data indicate that both of the these \ion{O}{6}
absorbers are located in large-scale filaments of the ``cosmic
web''.\label{sloanfig}}
\end{figure}

High-resolution UV spectra of PG0953+415 obtained with STIS and {\it
FUSE} have revealed two strong, intervening \ion{O}{6} absorbers in
the PG0953+415 spectrum at $z_{\rm abs}$ = 0.06807 and 0.14232 (Tripp
\& Savage 2000; Savage et al. 2002).  Some improvements in the data
reduction procedures, especially in the CALFUSE reduction software,
were implemented after publication of those papers, so we have
re-extracted the STIS and {\it FUSE} spectra as described in Tripp et
al. (2001, 2005).  The STIS spectrum has 7 km s$^{-1}$ resolution
(FWHM) and extends from 1150 to 1730 \AA , and the {\it FUSE} spectrum
covers $915 - 1187$ \AA\ at $20 - 25$ km s$^{-1}$ resolution.  The
STIS spectrum has an accurate wavelength scale zero point, and we
calibrated the {\it FUSE} wavelength scale zero point by aligning ISM
lines in the {\it FUSE} spectrum with comparable-strength lines in the
STIS spectrum (e.g., \ion{Fe}{2} $\lambda$1144.94 and \ion{Fe}{2}
$\lambda$1608.45). We remeasured absorption-line equivalent widths,
column densities, and Doppler parameters, and find good agreement with
the measurements reported in Savage et al. (2002).  In addition, for
the purposes of \S 3, we have constructed new apparent column density
profiles (Savage \& Sembach 1991) as follows. In a pixel at velocity
$v$, the apparent optical depth $\tau _{a} = {\rm ln}[I_{c}(v)/I(v)]$,
where $I(v)$ is the observed intensity and $I_{c}(v)$ is the estimated
continuum intensity, and the apparent column density $N_{\rm a}(v) =
(m_{\rm e}c/\pi e^{2})(f\lambda)^{-1}\tau _{\rm a}(v)$. The
\zabs{0.06807} system is detected in transitions of \ion{H}{1},
\ion{C}{4}, \ion{N}{5}, and \ion{O}{6}.  Comparison of the doublet
$N_{\rm a}(v)$ profiles indicates that the \ion{C}{4} and \ion{O}{6}
lines at \zabs{0.06807} are somewhat affected by saturation. For these
lines we use the $N_{\rm a}(v)$ profiles of the weaker line of the
doublet, which provides a conservative lower limit on the column.  The
\ion{O}{6} $N_{\rm a}(v)$ profiles at \zabs{0.14232} agree within the
noise and show no indications of unresolved saturation, so in this
case we further used the technique of Jenkins \& Peimbert (1997) to
construct weighted composite $N_{a}(v)$ profiles derived from both
members of the doublet.

To study the connections between galaxies and the PG0953+415
absorption systems, we have employed GMOS (Hook et al. 2004) on
Gemini-North to obtain galaxy redshifts.  Full details will be
published in a subsequent paper (B. Aracil et al., in prep).  In
brief, we first obtained a GMOS $5' \times 5'$ image centered on the
QSO in the SDSS r$'$ filter. The final image was constructed from
individual exposures that were dithered to fill in the chip gaps and
to allow cosmic ray rejection; the total exposure time was 38 minutes.
We used SExtractor (Bertin \& Arnouts 1996) to select
galaxies from the GMOS image for multislit follow-up spectroscopy with
the GMOS R150 grating and the GG455 order blocking filter; this setup
nominally covers the 4500 \AA\ - 1 $\mu$m wavelength range with a
resolving power of $\sim$600.  Targets were prioritized by brightness
and proximity to the QSO.

We obtained redshift measurements for 83 galaxies by cross correlation
with high signal-to-noise template spectra.  To graphically summarize
the redshift survey results, Figure~\ref{fig:gmos_img} (PLATE 1) shows
the GMOS image of the PG0953+415 field with the galaxies labeled with
their spectroscopic redshifts. Not surprisingly, most of the galaxies
are behind the QSO. We find only eight galaxies with redshifts at
least 5000 \kms\ lower than $z_{\rm QSO}$. These foreground galaxies
are indicated in white boxes in Figure~\ref{fig:gmos_img}. We note
that three of the foreground galaxies are near the redshift of the
\ion{O}{6} absorber at \zabs{0.14232} with the closest at a projected
distance\footnote{We assume $H_{0} = 70 h_{70}$ \kms Mpc$^{-1}$,
$\Omega _{\lambda}$ = 0.7, and $\Omega _{m}$ = 0.3 for the calculation
of projected distances in this paper.} $\rho = 155 h_{70}^{-1}$
kpc. To show the morphology of the three galaxies at $z \approx$
0.143, Figure~\ref{fig:gmos_zoom} (PLATE 2) presents a blow-up of the
NW quadrant of the GMOS image.  Figure~\ref{fig:gmos_zoom} is also
useful for closer inspection of galaxies close to the QSO.
Interestingly, the GMOS survey has not revealed any galaxies near the
\ion{O}{6} system at \zabs{0.06807}. Based on the SExtractor galaxy
catalog, we find that within 2.5$'$ of the quasar, our sample is 100\%
complete down to $r' = 19.7$, 88\% complete for $r' \leq 20.5$, and
73\% complete for $r' \leq 21.0$.  At $z=0.068$, $r' = 19.7$
corresponds to $L = 0.04\,L^*$. Savage et al. (2002)
have shown that the \zabs{0.06807} \ion{O}{6} absorber is associated
with a galaxy group or large-scale structure, but the nearest known
galaxy in that group has a large impact parameter, $\rho = 999
h_{70}^{-1}$ kpc. The new GMOS survey shows that if there is a galaxy
closer to the sight line at \zabs{0.06807}, it is either at $\rho \geq
195 h_{70}^{-1}$ kpc or is faint with $L \lesssim 0.04 L^{*}$.

PG0953+415 has also been covered by the {\it Sloan Digital Sky Survey}
({\it SDSS}, Adelman-McCarthy et al. 2006), including spectroscopic
follow-up.  Figure~\ref{sloanfig} shows the location of {\it SDSS}
galaxies at redshifts near the \zabs{0.06807} and 0.14232
absorbers. We only show galaxies within $\pm 600$ km s$^{-1}$ of each
absorber; galaxies with $z < z_{\rm abs}$ are shown with blue symbols,
and galaxies with $z > z_{\rm abs}$ are indicated with orange-red
symbols. From Figure~\ref{sloanfig}, it is immediately obvious that
both of these \ion{O}{6} systems are found within large-scale galaxy
filaments. Both filaments appear to feed into higher density regions
as expected in theoretical models (e.g., Dav\'{e} et
al. 1999). Inspection of the three-dimensional SDSS galaxy
distribution shows that these filaments are not merely coincidental
projection effects; these are real filaments and nodes of the ``cosmic
web''.  The closest galaxy in the {\it SDSS} sample at \zabs{0.06807}
has $\rho = 736 h_{70}^{-1}$ kpc. Note that the field of view shown in
Figure~\ref{sloanfig} is substantially larger than the field covered
by GMOS, but conversely our GMOS survey is much deeper than the {\it
SDSS}. The {\it SDSS} ``main galaxy'' spectroscopic sample targets
galaxies with $r <$ 17.8 (Strauss et al. 2002). This limiting
magnitude corresponds to $L \geq 0.2 L*$ and $L \geq 1.1 L*$ at
\zabs{0.06807} and \zabs{0.14232}, respectively.  The only galaxy in
the GMOS field for which {\it SDSS} provided a spectroscopic redshift
is the extended spiral galaxy in the upper right corner of
Figures~\ref{fig:gmos_img} and \ref{fig:gmos_zoom}.

\section{\ion{O}{6} Absorber Ionization and Abundances}\label{sec:abs}

\begin{figure}
\centering
\resizebox{1.00\hsize}{!}{\includegraphics{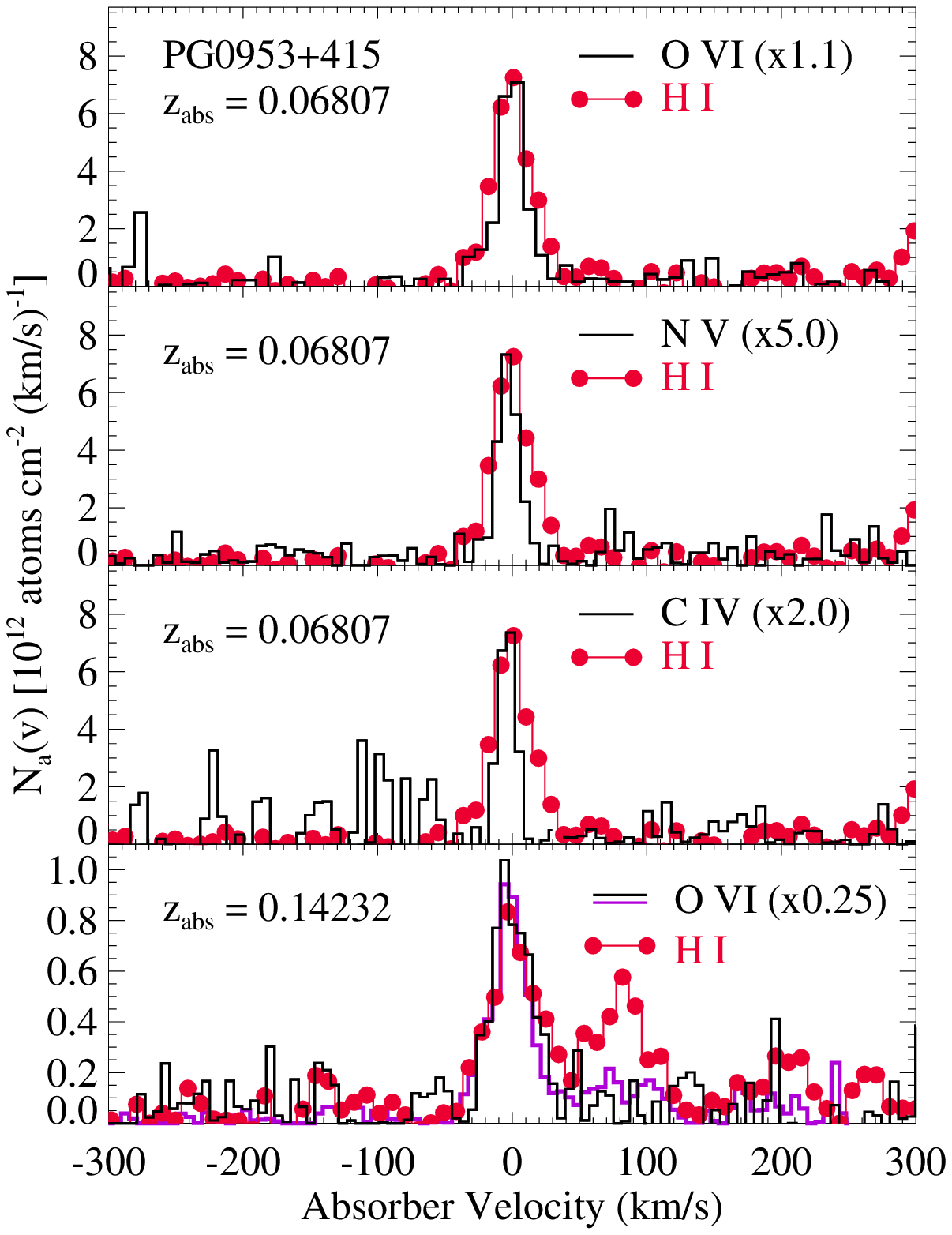}}
\caption{Apparent column density profiles of \ion{H}{1}, \ion{O}{6},
\ion{N}{5}, and \ion{C}{4} transitions observed at \zabs{0.06807}
({\it upper three panels}) and \ion{H}{1} and \ion{O}{6} at
\zabs{0.14232} ({\it lowest panel}) in the direction of PG~0953$+$415
. The profiles are plotted vs. absorber velocity (where $v = 0$ km
s$^{-1}$ at \zabs{0.06807} or \zabs{0.14232}), and the metal profiles
are scaled as indicated in each panel. In the lowest panel, the black
line shows the \ion{O}{6} profile derived from the STIS data; the
purple line indicates the \ion{O}{6} profile derived from the {\it
FUSE} spectra. The \zabs{0.06807} \ion{N}{5} and \zabs{0.14232}
\ion{O}{6} profiles are weighted composite profiles derived from both
lines of these doublets. The \zabs{0.06807} \ion{C}{4} and \ion{O}{6}
profiles are derived from the weaker line of the doublet, which
provide a lower limit on the column density. The \ion{H}{1} profile at
\zabs{0.06807} is based on the {\it FUSE} Ly$\beta$ profile, and the
\zabs{0.14232} \ion{H}{1} profile is derived from the STIS Ly$\alpha$
line.}\label{fig:nav}
\end{figure}

The centroids, velocity widths, and overall line shapes of the
\ion{H}{1} and metal absorption lines of the PG0953+415 \ion{O}{6}
systems are remarkably similar.  To show this, we overplot in
Figure~\ref{fig:nav} the apparent column density [$N_{a}(v)$] profiles
of the \ion{H}{1}, \ion{C}{4}, \ion{N}{5}, and \ion{O}{6} absorption
lines detected at \zabs{0.06807} ({\it upper three panels}) and the
\ion{H}{1} and \ion{O}{6} lines at \zabs{0.14232} ({\it lowest
panel}).  For this purpose, we have scaled the metal $N_{a}(v)$
profiles in Figure~\ref{fig:nav} to match the \ion{H}{1} profiles.

Figure~\ref{fig:nav} provides several valuable observational
constraints on the nature of these absorbers. Beginning with the
\zabs{0.06807} system, we see that the metal lines are precisely
aligned with the \ion{H}{1} absorption and the \ion{H}{1} profile is
narrow: $b$(\ion{H}{1}) = 21$\pm$2 \kms . Notably, {\it the \ion{H}{1}
lines are far too narrow to arise in collisionally ionized gas in
ionization equilibrium at the temperatures expected for \ion{O}{6}.}
The \ion{H}{1} line width at \zabs{0.06807} implies that $T \leq 3.8
\times 10^{4}$ K ($2\sigma $), i.e., an order of magnitude colder than
the temperature required to produce \ion{O}{6} in CIE. Moreover, the
close match of the \ion{H}{1} and the metal $N_{a}(v)$ profiles
strongly suggests that the \ion{H}{1} and metal absorption lines
originate in the same (single-phase) gas cloud.  The aborption profile
simplicity is another useful constraint. The \zabs{0.06807} system
shows mainly one narrow component, which suggests that the absorber is
a quiescent, undisturbed entity.

The \zabs{0.14232} system (bottom panel of Figure~\ref{fig:nav}) is
kinematically more complex: at this redshift, six components spread
over a velocity range of 450 \kms are present in the Ly$\alpha$
profile (see Fig. 6 in Tripp \& Savage 2000). Also, this case affords
fewer constraints on the gas ionization mechanism because \ion{C}{4}
and \ion{N}{5} are not detected. Nevertheless, it is interesting to
note that in the main, strongest component (at $v = 0$ \kms \ in
Figure~\ref{fig:nav}), the \ion{H}{1} and \ion{O}{6} profiles are
again seen to have very similar shapes, and again the \ion{H}{1} lines
are too narrow to arise in CIE [the main component at $v = 0$ has
$b$(\ion{H}{1}) $ = 28\pm 2$ km s$^{-1}$].

The striking similarity of the \ion{H}{1} and metal line profiles and
the narrow widths of the \ion{H}{1} lines are most easily understood
if the gas is photoionized by the UV background light from
QSOs/AGNs. In photoionized absorbers, the gas is expected to be cool
and quiescent.  In principle, these observations could alternatively
indicate that the absorbers are not in ionization equilibrium and have
cooled more rapidly than they can recombine. Or, these could be
multiphase absorbers, and the \ion{H}{1} absorption associated with
the hot \ion{O}{6} phase is not detected. However, currently available
non-equilibrium cooling gas models (Shapiro \& Moore 1976; Schmutzler
\& Tscharnuter 1993) require extraordinarily high metallicities ($Z
\geq 10 Z_{\odot}$) for agreement with the observed columns and line
widths in the PG0953+415 \ion{O}{6} systems. Based on these models,
non-eq. cooling gas appears to be highly unlikely.

Using CLOUDY (v96.01, Ferland et al. 1998), we have revisited the
properties predicted for the \zabs{0.06807} absorber if it is
photoionized.  We adopt the recent solar abundance revisions (see
Table 5 in Savage et al. 2005) and the modified Haardt \& Madau (1996)
UV background.\footnote{Motivated by Shull et al. (1999), the modified
UV background has a steeper EUV spectral index. Use of the original
Haardt \& Madau UV background with the shallower EUV index {\it
increases} the derived metallicity by a factor of $\sim$2. We
normalize the UV background to $J_{\nu} = 1 \times 10^{-23}$ ergs
s$^{-1}$ cm$^{-2}$ Hz$^{-1}$ sr$^{-1}$ at 1 Rydberg.}  We find that
our updated photoionization model is mostly in good agreement with the
results reported by Savage et al. (2002): we find that the logarithmic
gas metallicity [M/H] = $-0.3 \pm 0.12$, gas pressure $p/k \approx 1$
cm$^{-3}$ K, and absorber thickness = 70$^{+50}_{-30}$ kpc. In \S 2 we
noted that the \zabs{0.06807} absorption is apparently affected by
saturation, and therefore our \ion{O}{6} column could be
underestimated. Applying the method of Jenkins (1996) to correct the
\ion{O}{6} for saturation, we find that $N$(\ion{O}{6}) could be 0.25
dex higher than the value derived from profile fitting and direct
integration of the 1037.6 \AA\ line.\footnote{The \ion{C}{4} doublet
at \zabs{0.06807} also shows evidence of saturation, but the
\ion{C}{4} profiles are too noisy in the line cores to enable a
reliable saturation correction using the method of Jenkins (1996).}
This would seem to require an even higher metallicity.  However, such an increase in
\ion{O}{6} would require a higher ionization parameter (and therefore
a lower density and larger absorber size) in order to match the
resulting increased \ion{O}{6}/\ion{N}{5} ratio, and the required
change in metallicity would be small. If the \zabs{0.14232} absorber
is also photoionized, then the (albeit more limited) observations
again imply a low density and relatively high metallicity (see Savage
et al. 2002).

\section{Discussion}

The PG0953+415 \ion{O}{6} absorbers appear to be photoionized and
metal-rich, and yet the nearest galaxies are relatively far from the
sight line.  Other recent studies of absorber-galaxy connections have
found similar results, i.e., high-metallicity, highly ionized gas in
large-scale structures but with no luminous galaxies within $\rho
\lesssim$ 100 kpc (e.g., Tumlinson et al. 2005; Aracil et al. 2006;
Prochaska et al. 2006). The detection of enriched, photoionized
\ion{O}{6} absorbing gas is in contrast to the predictions that IGM
\ion{O}{6} absorbers arise in collisionally-ionized metal-poor WHIM
gas. Our detection of kinematically simple photoionized gas so far
from any galaxy also does not conform to the idea that IGM O~VI arises
predominantly in turbulent collisionally-ionized galactic
outflows. How, then, did such chemically enriched gas end up more than
150 kpc away from any known luminous galaxy?

One possibility is that the \ion{O}{6} absorbers arise in material
that has been dynamically stripped out of a galaxy.  It has long been
known that tidally stripped gaseous structures can extend for many
tens of kpc (e.g., Yun, Ho, \& Lo 1994), and \ion{O}{6} absorption is
frequently detected in sight lines through the Magellanic Stream
(Sembach et al. 2003), a nearby example of a dynamically stripped
material.  This hypothesis can explain the lack of a nearby luminous
galaxy because the donor galaxy could be a faint satellite dwarf which
would be hard to detect. The relatively high metallicity of the
absorbers can also be understood because tidally stripped material can
be significantly enriched (reflecting the metallicity of the
progenitor galaxy). Tidal debris is often kinematically complex (e.g.,
Bowen et al. 1994), and while the \zabs{0.14232} system shows
similarly complex kinematics, the \zabs{0.06807} absorber does not.
In addition, both PG0953+415 \ion{O}{6} systems have much lower
\ion{H}{1} column densities than well-studied tidal structures.
However, {\it FUSE} observations have revealed absorption at
Magellanic Stream velocities without corresponding 21 cm emission
(Sembach et al. 2003); much tidal detritus is likely to have
\ion{H}{1} columns below the 21 cm detection threshold.  It remains
possible that the absorption arises in the lower-density, more
quiescent periphery of a tidal structure. A more substantial problem
is that \ion{O}{6} in the Magellanic Stream is believed to be {\it
collisionally} ionized due to interaction between the stream and a hot
Galactic halo, and the \ion{O}{6} and low ions in the stream often
show different velocity centroids and widths, unlike the PG0953+415
\ion{O}{6} systems.  However, some Magellanic Stream clouds are
kinematically simple and even show \ion{C}{1} and H$_{2}$ absorption
lines (Sembach et al. 2001), so tidal stripping can be a gentle
process for at least some of the stripped gas.  Once the density of
the tidal debris drops sufficiently, it will become highly ionized by
UV background photoionization.

However, galaxies are known to show a correlation between metallicity
and luminosity (Tremonti et al. 2004, and references therein), so if
the PG0953+415 \ion{O}{6} systems were stripped out of a
low-luminosity dwarf, one might expect the absorbers to have a lower
metallicity. An alternative hypothesis is that the PG0953+415
\ion{O}{6} absorbers might have been mainly enriched by processes that
occurred at high redshifts.  Recent studies indicate that
star-forming, high$-z$ galaxies can have near-solar metallicities at
$z > 2$ (e.g., Pettini et al. 2002; Shapley et al. 2004; Erb et
al. 2006), and Simcoe et al. (2006) have reported evidence that QSO
absorbers have high metallicities within 100-200 $h_{70}^{-1}$ kpc of
high$-z$ galaxies. This hypothesis can explain the absence of a nearby
galaxy in the case of the \zabs{0.06807} absorber toward PG0953+415:
if the enrichment occurred at $z \gtrsim 2$, then the source galaxy
could have moved a substantial distance away from the sight line
during the time that has passed since the metals were injected into
the IGM.  In this hypothesis, the metals could have been removed from
the source galaxy by a supernova-driven outflow or by dynamical
stripping; in either case the enriched gas could settle into a
quiescent state and thereafter remain highly ionized due to
photoionization by the UV background.  To test these ideas, it would
be useful to measure the metallicities of the nearest galaxies.  The
three galaxies closest to the \zabs{0.14232} absorber (see
Figure~\ref{fig:gmos_zoom}) are all emission-line galaxies, as are
some of the closest galaxies at \zabs{0.06807} (see Aracil et al., in
prep).  Comparison of the galaxy and absorber metallicities would
provide additional constraints on the possible origin of the
enrichment.

%
We thank Kathy Roth for valuable assistence with the setup of the GMOS
observations for the Gemini queue, and we thank Mauro Giavalisco for
interesting comments on this paper. TMT appreciates support from NASA
grant NNG-04GG73G, and DVB acknowledges NASA grant
NNG-05GE26G. Funding for the SDSS and SDSS-II has been provided by the
Alfred P. Sloan Foundation, the Participating Institutions, the
National Science Foundation, the U.S. Department of Energy, the
National Aeronautics and Space Administration, the Japanese
Monbukagakusho, the Max Planck Society, and the Higher Education
Funding Council for England. The SDSS Web Site is
http://www.sdss.org/.


\begin{thebibliography}{}

\bibitem[Adelman-McCarthy et al. (2006)]{dr4} Adelman-McCarthy,
J. K. et al. 2006, \apjs, 162, 38

\bibitem[Aracil et al. (2006)]{aracil05} Aracil, B., Tripp, T. M.,
Bowen, D. V., Prochaska, J. X., Chen, H.-W., \& Frye, B. L. 2005,
\mnras, 367, 139

\bibitem[Bertin \& Arnouts (1996)]{bertin96} Bertin, E., \& Arnouts,
S. 1996, A\&AS, 117, 393

\bibitem[Bowen et al. (1994)]{bowen94} Bowen, D. V., Roth, K. C.,
Blades, J. C., \& Meyer, D. M. 1994, \apj, 420, L71

\bibitem[Bregman et al. (2004)]{breg04} Bregman, J. N., Dupke, R. A.,
\& Miller, E. D. 2004, \apj, 614, 31

\bibitem[Cen \& Ostriker (1999)]{cen99} Cen, R., \& Ostriker,
J. P. 1999, \apj, 514, 1

\bibitem[Danforth \& Shull (2005)]{dan05} Danforth, C. W., \& Shull,
J. M. 2005, \apj, 624, 555

\bibitem[Dav\'{e} et al. (1999)]{dave99} Dav\'{e}, R., Hernquist, L.,
Katz, N., \& Weinberg, D. H. 1999, \apj, 511, 521

\bibitem[Erb et al. (2006)]{erb06} Erb, D. K., Shapley, A. E.,
Pettini, M., Steidel, C. C., Reddy, N. A., \& Adelberger, K. L. 2006,
\apj, in press (astro-ph/0602473)

\bibitem[Ferland et al. (1998)]{ferland98} Ferland, G. J., Korista,
K. T., Verner, D. A., Ferguson, J. W., Kingdon, J. B., \& Verner,
E. M. 1998, \pasp, 110, 761

\bibitem[Haardt \& Madau (1996)]{hm96} Haardt, F., \& Madau, P. 1996,
\apj, 461, 20

\bibitem[Hook et al. (2004)]{hook04} Hook, I. et al. 2004, \pasp, 116,
425

\bibitem[Jenkins (1996)]{jenkins96} Jenkins, E. B. 1996, \apj, 471,
292

\bibitem[Jenkins \& Peimbert (1997)]{jp97} Jenkins, E. B., \&
Peimbert, A. 1997, \apj, 477, 265

\bibitem[Lehner et al. (2006)]{lehner06} Lehner, N., Savage, B. D.,
Wakker, B. P., Sembach, K. R., \& Tripp, T. M. 2006, \apjs, in press
(astro-ph/0602085)

\bibitem[Pettini et al. (2002)]{pettini02} Pettini, M., Rix, S. A.,
Steidel, C. C., Adelberger, K. L., Hunt, M. P., \& Shapley,
A. E. 2002, \apj, 569, 742

\bibitem[Prochaska et al. (2005)]{prochaska05} Prochaska, J. X.,
Weiner, B. J., Chen, H.-W., \& Mulchaey, J. S. 2006, \apj, in press
(astro-ph/0602172)

\bibitem[Savage et al. (2005)]{Savage05} Savage, B. D., Lehner, N.,
Wakker, B. P., Sembach, K. R., \& Tripp, T. M. 2005, \apj, 626, 776

\bibitem[Savage \& Sembach (1991)]{savage91} Savage, B. D., \&
Sembach, K. R. 1991, \apj, 379, 245

\bibitem[Savage et al. (2002)]{Savage02} Savage, B. D., Sembach,
K. R., Tripp, T. M., \& Richter, P. 2002, \apj, 564, 631

\bibitem[Schmutzler \& Tscharnuter (1993)]{schmutz} Schmutzler, T. \&
Tscharnuter, W. M. 1993, \aa, 273, 318

\bibitem[Sembach et al. (2003)]{sembach03} Sembach, K. R., et
al. 2003, \apjs, 146, 165

\bibitem[Sembach et al. (2001)]{sembach01} Sembach, K. R., Howk,
J. C., Savage, B. D., \& Shull, J. M. 2001, \aj, 121, 992

\bibitem[Sembach et al. (2004)]{sembach04} Sembach, K. R., Tripp,
T. M., Savage, B. D., \& Richter, P. 2004, \apjs, 155, 351

\bibitem[Shapiro \& Moore (1976)]{shapmore} Shapiro, P. R., \& Moore,
R. T. 1976, \apj, 207, 460

\bibitem[Shapley et al. (2004)]{shap04} Shapley, A. E., Erb, D. K.,
Pettini, M., Steidel, C. C., \& Adelberger, K. L. 2004, \apj, 612, 108

\bibitem[Shull et al. (1999)]{shull99} Shull, J. M., Roberts, D.,
Giroux, M. L., Penton, S. V., \& Fardal, M. A. 1999, \aj, 118, 1450

\bibitem[Simcoe et al. (2006)]{simcoe06} Simcoe, R. A., Sargent,
W. L. W., Rauch, M., \& Becker, G. 2006, \apj, 637, 648

\bibitem[Stocke et al. (2005)]{stocke05} Stocke, J. T., Penton, S. V.,
Danforth, C. W., Shull, J. M., Tumlinson, J., \& McLin, K. M. 2005,
\apj, 641, 217

\bibitem[Strauss et al. (2002)]{strauss02} Strauss, M. A. et al. 2002,
\aj, 124, 1810

\bibitem[Tremonti et al. (2004)]{tre04} Tremonti, C. A. et al. 2004,
\apj, 613, 898

\bibitem[Tripp et al. (2001)]{Tripp01} Tripp, T. M., Giroux, M. L.,
Stocke, J. T., Tumlinson, J., \& Oegerle, W. R. 2001, \apj, 563, 724

\bibitem[Tripp et al. (2005)]{Tripp05} {Tripp}, T.~M., {Jenkins},
E.~B., {Bowen}, D.~V., {Prochaska}, J.~X., {Aracil}, B. \& {Ganguly},
R. 2005, {\apj}, 619, 714

\bibitem[Tripp \& Savage (2000)]{ts00} Tripp, T. M., \& Savage,
B. D. 2000, \apj, 542, 42

\bibitem[Tripp et al. (2000)]{tripp00} Tripp, T. M., Savage, B. D., \&
Jenkins, E. B. 2000, \apj, 534, L1

\bibitem[Tumlinson et al. (2005)]{tum05} Tumlinson, J., Shull, J. M.,
Giroux, M. L., \& Stocke, J. T. 2005, \apj, 620, 95

\bibitem[Voit (2005)]{voit05} Voit, G. M. 2005, Rev. Mod. Physics, 77,
207

\bibitem[Yun et al. (1994)]{yun94} Yun, M. S., Ho, P. T. P., \& Lo,
K. Y. 1994, Nature, 372, 530

\end{thebibliography}
\end{document}